\documentclass[journal=jacsat,manuscript=article,layout=twocolumn]{achemso}
\usepackage{graphicx}
\usepackage{multicol}
\usepackage{parskip}
\usepackage{enumitem}
\usepackage{siunitx} 
\usepackage[version=3]{mhchem} 
\usepackage{xcolor}

\usepackage{lineno}

\author{Jelena Wohlwend}
\email{jelena.wohlwend@mat.ethz.ch}
 \affiliation{Laboratory for Nanometallurgy, Department of Materials, ETH Zurich, 8093 Zürich, Switzerland}
\author{Oliver Wipf}
 \affiliation{Laboratory for Nanometallurgy, Department of Materials, ETH Zurich, 8093 Zürich, Switzerland}
   \author{David Kiwic}%
 \affiliation{Laboratory for Multifunctional Materials, Department of Materials, ETH Zurich, 8093 Zürich, Switzerland}
 \author{Siro Käch}%
 \affiliation{Laboratory for Nanometallurgy, Department of Materials, ETH Zurich, 8093 Zürich, Switzerland}
 \author{ Benjamin Mächler}
 \affiliation{Laboratory for Nanometallurgy, Department of Materials, ETH Zurich, 8093 Zürich, Switzerland}
 \author{Georg Haberfehlner}%
 \affiliation{Institut für Elektronenmikroskopie und Nanoanalytik, TU Graz, 8010 Graz, Austria}
 \author{Ralph Spolenak}%
 \affiliation{Laboratory for Nanometallurgy, Department of Materials, ETH Zurich, 8093 Zürich, Switzerland}
 \author{Henning Galinski}%
 \affiliation{Laboratory for Nanometallurgy, Department of Materials, ETH Zurich, 8093 Zürich, Switzerland}
\title[An \textsf{achemso} demo]
  {\textbf{CO$_2$ Conversion in Cu-Pd based Disordered Network Metamaterials with Ultra-Small Mode Volumes}}

\abbreviations{IR,visible}

\begin{document}

\begin{abstract}
Plasmons can drive chemical reactions by directly exciting intramolecular transitions. However, strong coupling of plasmons to single molecules remains a challenge as ultra-small mode volumes are required. In the presented work, we propose Cu-Pd plasmonic network metamaterials as a scalable platform for plasmon-assisted catalysis. Due to the absence of translational symmetry, these networks provide a unique plasmonic environment featuring a large local density of optical states and an unparalleled density of hotspots that effectively localizes light in mode volumes $V<8\cdot10^{-24}$~m$^3$. Catalytic performance tests during CO$_2$ conversion reveal production rates of up to 4.3$\cdot$10$^2$ mmol g$^{-1}$h$^{-1}$ and altered reaction selectivity under light illumination. Importantly, we show that the selectivity of the catalytic process can be tuned by modifying the network’s chemical composition, offering a versatile approach to optimize reaction pathways. 
\end{abstract}
\section{Letter}
\textbf{plasmonics, disordered photonics, self-assembled metamaterials, CO$_2$ conversion, local density of optical states, EELS}
\par 
Reducing CO$_2$ in our atmosphere is a necessity in order to avert the pending climate crisis caused by the continuous emission of CO$_2$ through the burning of fossil fuels like coal, oil and natural gas. For this, new technologies have to be developed to capture and store CO$_2$, and possibly convert CO$_2$ into new functional chemicals, such as e-fuels. The conversion of CO$_2$ into e-fuels is especially intriguing as it offers a sustainable solution to close the carbon cycle. 
\par
E-fuel synthesis is commonly achieved in specifically designed antenna-reactor complexes, where non-radiative plasmonic decay is utilized to drive chemical reactions by light~\cite{bimetallicnanostructuresplasmonicsandcatalysisdionne,antennareactor}. Plasmon-assisted catalysis is a synergistic effect, and the role and interaction between the single contributions are the subject of intense scientific debates~\cite{verma2024paradox,doi:10.1126/science.aao0872,doi:10.1021/acs.jpcc.0c08831,doi:10.1021/acs.nanolett.7b00992,limitsofhotcarrierinjection}. Mechanisms at play include near-field enhancement, local heating and "hot" carrier generation~\cite{hotelectronsandthermaleffects,Plasmonassistedco2reduction,hotelectrongeneration}. While near-field enhancement and local heating are classical phenomena the generation of "hot" carriers is a quantum effect. It results from Landau damping, i.e. the non-conservation of linear momentum of electrons near surfaces and regions of high field enhancement (hotspots). Most conventional plasmonic antenna-reactor designs are based on two-dimensional systems, such as metasurfaces~\cite{dongare20213d, eelsjeniffer, mascaretti2022challenges, yuan2023quasi}, dispersed nanoparticles~\cite{antennareactor,herran2023plasmonic,kang2024effect}, and nanosheets~\cite{nonmetalliccatalyst,wang2024plasmonic}. However, the low surface-to-volume ratio in such 2D systems impairs on their performance. It limits both, the total number of reaction sites on such surfaces and the hotspot density. This constraint is critical under realistic illumination conditions as only one plasmon polariton per nanoparticle exists at a given time and the time between excitations can be as long as microseconds~\cite{limitsofhotcarrierinjection}.
\begin{figure*}[t!]
\includegraphics[width=1.0\textwidth]{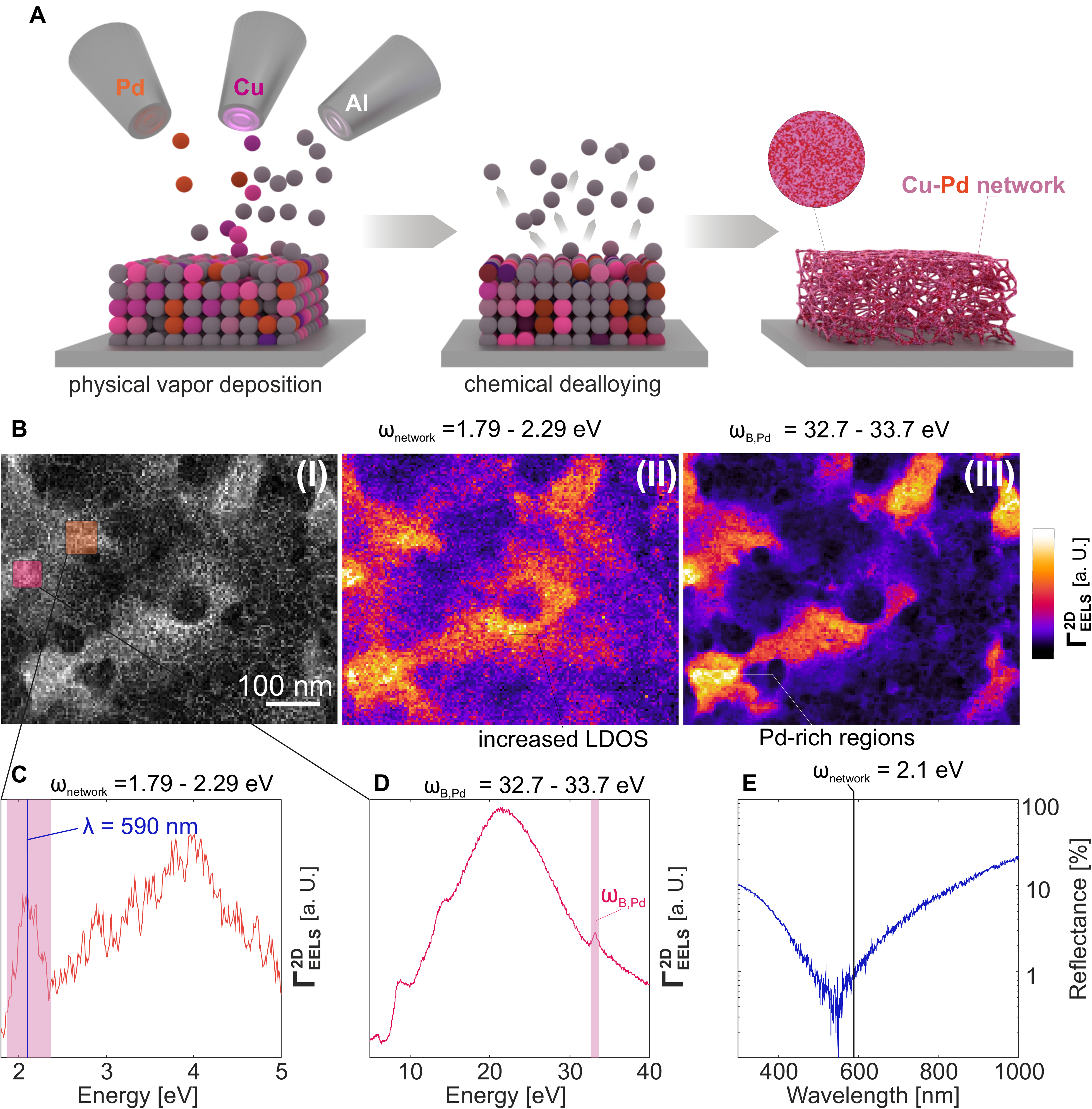}
\caption{\label{fig:Catalysis_one} \textbf{Fabrication and characterization of Cu-Pd disordered network metamaterials.} \textbf{A} Fabrication of the disordered network metamaterials by a two-step process: physical vapor deposition followed by chemical dealloying. \textbf{B} (I) HAADF micrographs of a selected Cu-Pd disordered network metamaterial (II) color-coded maps of global network mode (III) and bulk plasmon excitation. EEL spectra of the global network mode \textbf{C}  and bulk mode \textbf{D} of the network taken from "hot" spots. The regions are indicated with colored squares in the HAADF micrograph. \textbf{C} Localized plasmonic eigenmodes in a selected "hot" spot. The energy range corresponding to eigenmodes at visible frequencies $\omega_{network}$ is shaded in pink. \textbf{D} High energy EEL spectrum in a selected "hot" spot showing the bulk plasmon peak $\omega_{B,Pb}$ of Pd (shaded in pink). \textbf{E} Near-normal incidence reflectance spectrum of the Cu-Pd network (far field response).}
\end{figure*}
The confinement of the catalytic reactions to a plane also promotes local depletion effects of the reactants, leading to poor conversion performance \cite{gasdepletion}. Furthermore, simple geometries such as discs or particles only allow for moderate localization of light and are inherently bandwidth limited~\cite{Metamaterialswithindexellipsoids_Pendry}. This further constrains the injection of "hot" carriers~\cite{limitsofhotcarrierinjection}. The combination of large mode volumes $V$ and low quality factors $Q$~\cite{naya2020ag,yuan2023quasi,bi2022strong} of the resonant elements aggravates strong coupling in these two-dimensional systems~\cite{bitton2022plasmonic,lee2023strong}.
\begin{figure*}[t!]
\includegraphics[width=1.0\textwidth]{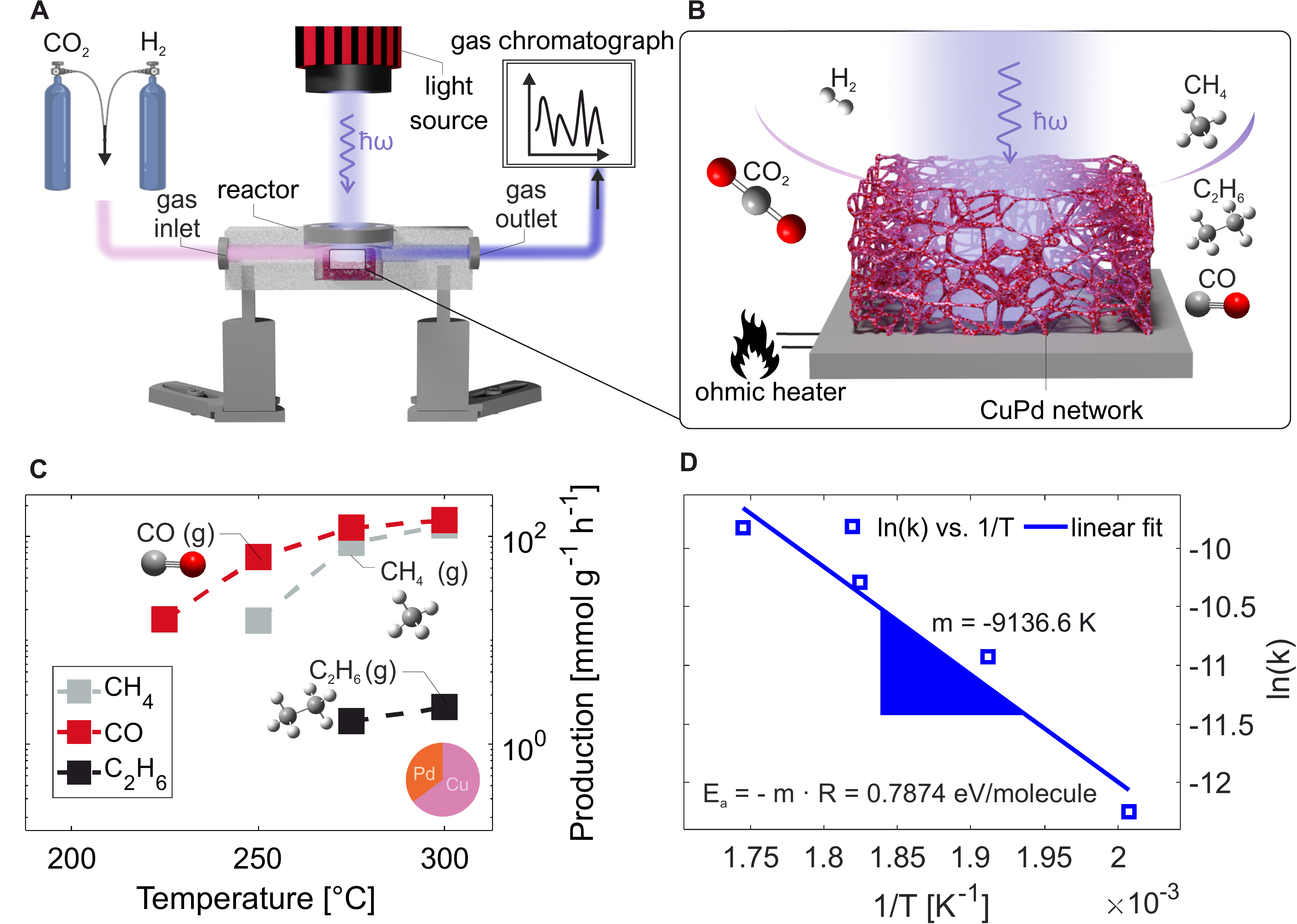}
\caption{\label{fig:Catalysis_two} \textbf{Temperature-dependent catalytic conversion of CO$_2$.} \textbf{A} and \textbf{B} Schematic illustration of the custom-built reactor, reactants and products, with gas inlet on the left and gas outlet on the right leading to the gas chromatograph. The chamber is sealed by a quartz window and a high-temperature O-ring, enabling illumination of the sample with a LED light source (white or UV=365~nm). The reactor can be heated by an ohmic heater (T$_\text{max}=300^\circ$C). \textbf{C} Temperature dependent production of \ce{CO}, \ce{CH4}, \ce{C2H6} without optical illumination with a 65\%Cu 35\%Pd network. \textbf{D} Arrhenius plot of the catalytic conversion of \ce{CO2} to \ce{CO}. Calculation of the activation energy E$_a= \SI{0.7874}{eV/molecule}$ for the conversion of CO$_2$ to CO using the reaction constant k, calculated from the temperature dependent yield.}
\end{figure*}
\par
In this work, we propose disordered network metamaterials (DNM) as a scalable platform for plasmonic catalysis in the strong coupling regime. Unlike two-dimensional antenna-reactor systems, these self-assembled structures combine a high surface-to-volume ratio and high density of hotspots that effectively localize light in ultra-small mode volumes $V$. 
Such small mode volumes $V$ are expected to enhance the dipolar plasmon-molecule interaction, as its coupling strength scales with $\Omega \sim \frac{1}{\sqrt{V}}$~\cite{Yoo2021}. 
In particular, we use Cu-Pd based plasmonic networks to show the successful design of scalable light harvesting catalysts. These DNMs enable quasi-perfect absorption of light, selectivity for specific reaction pathways and plasmon-enhanced catalytic conversion. DNMs are fabricated by a simple two-step process: physical vapor deposition of a ternary alloy and self-assembly by chemical dealloying (Figure \ref{fig:Catalysis_one}~\textbf{A})~\cite{lightmanipulationinmetallicnano,wohlwend_chemicalengineering,wohlwend2024hybrid}. 
\par 
The catalytic performance of the DNMs is screened using the light-assisted conversion of carbon dioxide at temperatures between $225$~$^\circ$C and $300$~$^\circ$C. Bimetallic Cu-Pd systems have been shown to be effective catalysts for CO$_2$ hydrogenation reactions~\cite{selectivehydrogenationofCO2,bimetallicpdcucatalysts}. The advantage of our systems lies in the capacity to modify the local chemistry and architecture of disordered networks easily during the first fabrication step. This provides a powerful approach to tailor both selectivity and absorption strength by adjusting the position of the d-band centers (see Supporting Information)~\cite{highselectivity,bimetallicnanostructuresplasmonicsandcatalysisdionne}.
\par 
To study the localization of light in our networks we employ energy electron loss spectroscopy (EELS). EELS has emerged as the ideal technique to characterize plasmonic structures on the nanoscale. EELS maps do not only offer insights into local plasmonic excitations but also provide detailed insights into nanoscale chemistry. For ultra-thin systems, the low energy EELS signal $\Gamma_{\text{EELS}}^{2D}$ is directly proportional to the local density of optical states $\rho$ (LDOS), while the high energy signal correlates with the electron density~\cite{kociakmappingplsamonsnanometerscale, HaberfehlnerEELSphotonicenviroment, modecouplingheterodimer, aluminumcayleytrees,wohlwend2023} (Figure \ref{fig:Catalysis_one}~\textbf{C,D}). The LDOS describes the electromagnetic environment of a dipolar emitter and critically impacts the light-matter interaction, e.g. intermolecular transitions in a plasmonic cavity~\cite{fluctuationsintheLDOS}. 
\par 
In Figure \ref{fig:Catalysis_one} \textbf{B} (I) a high angle annular dark field (HAADF) micrograph with corresponding EEL maps of a Cu-Pd network, with a Pd content of 35 vol. \%  is shown. The HAADF micrograph confirms the formation of a nanosized network. 
\par 
To visualize the distribution of the plasmonic modes, we integrate the EELS signal within a spectral window from 1.79 to 2.29 eV (Figure~\ref{fig:Catalysis_one} \textbf{B} (II)). We observe large fluctuations in the LDOS modulated within the network. Such fluctuations confirm both an increased contribution of localized plasmonic eigenmodes and their ability to localize energy at nanometric scales in so-called ”hotspots” \cite{fluctuationsintheLDOS}. Measurements of the spatial extension of these modes reveal typical mode volumes of $20 \times 20 \times 20$~nm$^3$. By analyzing the bulk plasmon peak (Figure~\ref{fig:Catalysis_one}~\textbf{B} (III)) we can link these fluctuations in LDOS to modulations in the Pd content. 
\par 
Within the hotspots we observe a bimodal spectrum of plasmonic modes. One set of modes is centered at 2.1 eV, corresponding to visible wavelengths ($\approx$590~nm) and the other set of modes centered at 3.95~eV corresponding to ultra violet wavelength ($\approx$313~nm)(Figure~\ref{fig:Catalysis_one}~\textbf{C}).
Interestingly, the plasmonic modes can also be accessed by far-field techniques such as reflectance measurements. The far-field optical response, i.e. the reflectance spectrum in Figure~\ref{fig:Catalysis_one}~\textbf{E} corresponds well to the plasmonic eigenmodes in the visible regime in Figure~\ref{fig:Catalysis_one}~\textbf{C}. We observe quasi-perfect absorption over a broad range of wavelength. The slight deviation most probably stems from the difference in excitation (photons or electrons).
\par 
In regions of high LDOS and small mode volumes $V$, i.e. high near-field enhancement, the potential energy surface of catalytic reaction is prone to change. Two specific scenarios are of interest, namely the reduction of the activation energy $E_{a,j}$ and the change in selectivity if different products are formed. Especially the change in selectivity implicates non-thermal effects as the total power $Q$ dissipated in the metamaterial due to plasmon-induced heating is not reaction specific and proportional to the local optical losses $Q\sim Im(\epsilon)$~\cite{heatgenerationinplasmonicnanostructures}.
It is to note, that a high LDOS and high optical loss should generally enhance the reaction rate as the temperature is locally increased.
\begin{figure*}[t!]
\includegraphics[width=1\textwidth]{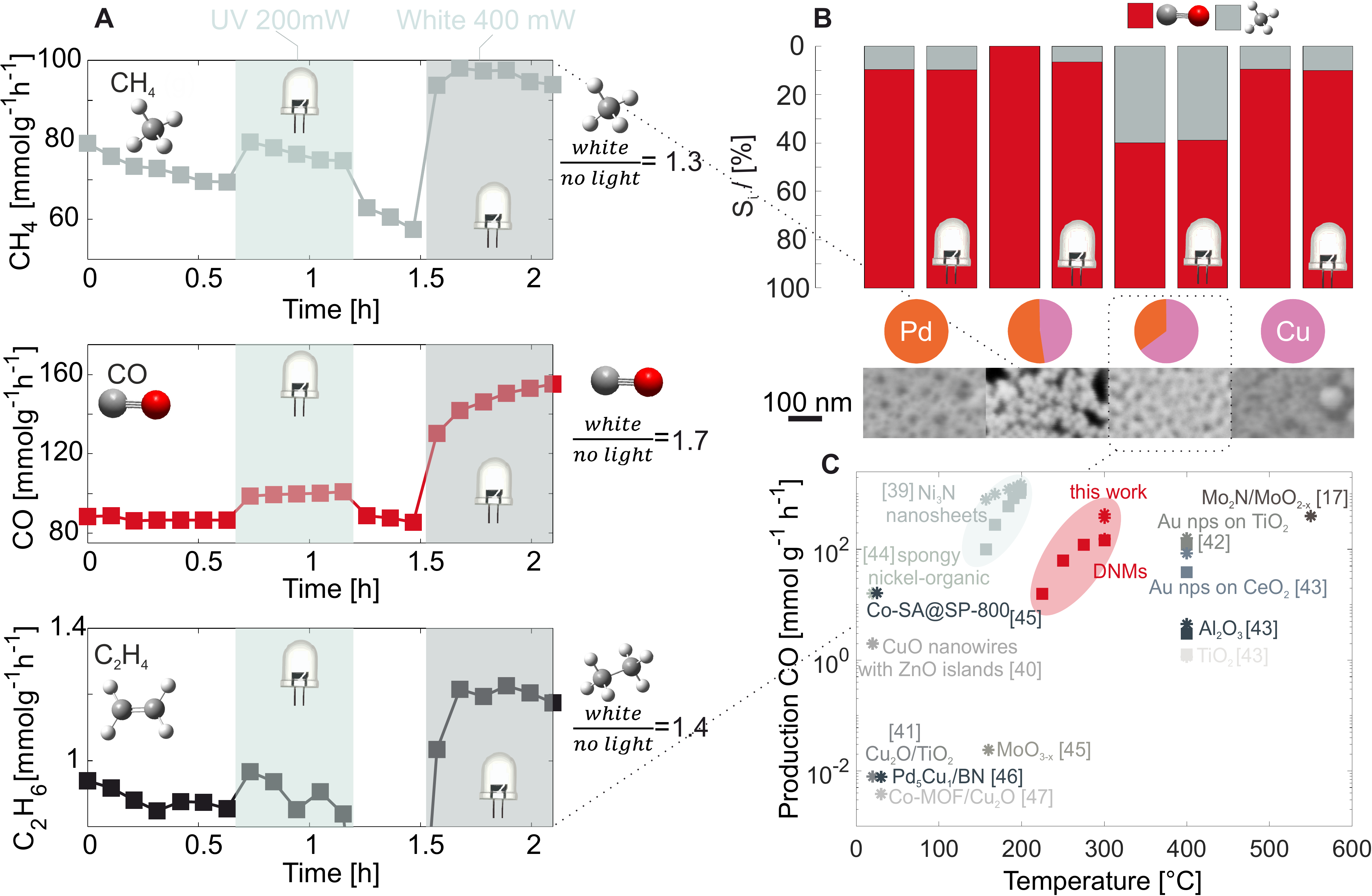}
\caption{\label{fig:Catalysis_three} \textbf{"Light-assisted" CO$_2$ conversion.} \textbf{A} "Light-assisted" \ce{CO2} conversion at constant temperature (T$_\text{max}=300^\circ$C) as a function of time. The conversion enhancement with white LED illumination (vs. no light) is depicted on the right. Especially, the \ce{C2H6} production shows a clear selectivity increase between UV irradiation (200 mW) and white light irradiation (400 mW). \textbf{B} Composition dependent selectivity $S_i$ for disordered network metamaterials with different Pd/Cu content and the corresponding scanning electron micrographs. \textbf{C} Comparison of the temperature dependent CO production of Cu-Pd networks with state-of-the-art catalyst from literature~\cite{nanosheets,CuOnanowireswithZNOIslands,Cu2OTiO2,Autio2,oxidesupportedaucatalysts,overviewcatalysisco2,nonmetalliccatalyst,photothermalcatalistmo03x,BNPDcucatalysis,comofcu2o}  }
\end{figure*}
\par 
To study the impact of the plasmonic environment of our DNMs on their catalytic performance, we carried out CO$_2$ conversion experiments in a custom build reactor (Figure \ref{fig:Catalysis_two}~\textbf{A}). Thereby, the sample is placed into a reaction chamber which is sealed by a quartz glass window and heated up (between 225 °C and 300 °C) while the chamber is flushed with CO$_2$ and H$_2$. The gas mixture escaping the reactor is directed into a gas chromatograph, which identifies the reaction products based on their distinct retention times.
\par 
During the catalytic conversion of CO$_2$, three main types of products are expected: CO, methanol, and hydrocarbons~\cite{catalyticconversionCO2} (Figure \ref{fig:Catalysis_two} \textbf{B}). CO can be produced via a reverse water-gas shift (RWGS) reaction, an endothermic process in which CO$_2$ is hydrogenated to produce syngas (H$_2$ and CO) and water, as shown in Equation \ref{catalysis_co2_1}.
\begin{equation}
\label{catalysis_co2_1}
CO_2 + H_2 \leftrightarrow CO + H_2O 
\end{equation}
The resulting syngas can be further utilized to synthesize methanol or as feedstock in the Fischer-Tropsch reaction to produce e-fuels~\cite{fishertrop}. In addition to CO various hydrocarbons can also be synthesized. For example, CH$_4$ (methane) is desirable as an e-fuel or as feedstock for producing synthetic natural gas (SNG). Longer-chain hydrocarbons, such as C$_2$H$_6$ (ethane), are valuable building blocks for further chemical synthesis or as components in fuels. Beyond the reaction products, the selectivity of the catalyst is also of great importance as it allows us to choose a product. If a catalyst provides high selectivity, tedious purification processes can be avoided.
\par 
In first experiments without illumination, the temperature was gradually increased from 150 °C to 300 °C in steps of 25 °C (Figure \ref{fig:Catalysis_two} \textbf{C}). At an onset temperature of 225 °C, the CO production starts, whereas the CH$_4$ production only starts at slightly higher temperatures of 250 °C. The production of ethane (C$_2$H$_6$) starts at 275 °C. At 300 °C the network converts CO$_2$ into CO, CH$_4$ and C$_2$H$_6$ with a selectivity of 60\% for CO, 39\% for CH$_4$ and 1\% for C$_2$H$_6$. The activation energy can be derived from the variation of yield with temperature (for details on the calculation see Supporting Information). The calculated activation energy for the reduction of CO$_2$ to CO is E$_a= \SI{0.7874}{eV/molecule}$ (Figure \ref{fig:Catalysis_three}~\textbf{C}) and comparable with findings by Bossche \textit{et al.}~\cite{activationenergyco2bosche} and Hussain \textit{et al.}~\cite{activationenergyco2hussain} for the electroreduction of CO$_2$.
\par 
In a second set of experiments, we evaluate the influence of light on the catalytic conversion of CO$_2$ at a constant temperature of 300 °C by illuminating the metamaterial with two different light sources: a 200 mW/cm$^2$ LED at 365 nm and 400 mW/cm$^2$ white LED (400-800) nm (Figure \ref{fig:Catalysis_three}~\textbf{A}). The production of all three reaction products is significantly enhanced by light. Specifically the maximal enhancement is 1.3 for CH$_4$, 1.7 for CO and 1.4 for C$_2$H$_6$ with white light illumination (Figure~\ref{fig:Catalysis_three}~\textbf{A}). Notably, the ethane production exhibits a distinct increase in selectivity between UV and white light. While UV illumination yields no significant increase in production, the presence of white light enhances the ethane production by a factor of 1.4. This finding is important as it suggests a frequency dependence of the process, likely correlating with the plasmonic response measured by EELS  (resonance at 2.1 eV $= 590$~nm).
\par 
Furthermore, the influence of light on the selectivity also depends on the network compositions (Figure \ref{fig:Catalysis_three}~\textbf{B}). 
Both chemical composition as well as network architecture significantly influence selectivity. For the pure Pd network, no increase in production and no change in selectivity is observed (see Supporting Information), whereas for the network with additional Cu content, an increase in production as well as a change in product selectivity is observed (Figure \ref{fig:Catalysis_three}~\textbf{B}). Especially, the 48\% Cu 52\% Pd networks shows a distinct selectivity change from only CO to additional CH$_4$ production with the illumination of light. 
\par
This change in selectivity indicates a plasmonic contribution to the catalytic conversion of CO$_2$  as it implicates a non-thermal frequency selective effect. Yet, it remains to be determined if this plasmon induced enhancement result from field enhancement or "hot" carrier generation. 
\par 
In order to compare the measured catalytic performance of our plasmonic networks with literature, we calculate the weight $g_{cat}$ of the DNMs using focused-ion-beam tomography (see Supporting Information).
\par 
Cu-Pd disordered network metamaterials exhibit catalytic performances ranging from 1.46 to \SI{98.15}{mmol \cdot g_{cat}^{-1} h^{-1}} for
the production of \ce{CH4} and from 16.49 to \SI{430.53}{mmol \cdot g_{cat}^{-1} h^{-1}} for the production of \ce{CO} across the network compositions. 
Notably, the catalytic performance of our plasmonic Cu-Pd based networks is on par with state-of-the-art high performing catalysts, such as Mo$_2N/MoO_{2-x}$~\cite{nonmetalliccatalyst} or Ni$_3$N nanosheets~\cite{nanosheets} (see Figure~\ref{fig:Catalysis_three}~\textbf{C}). We attribute this high performance of our DNMs to the ultra-small mode volumes which are a direct result of the self-assembly process that determines the architecture and chemical modulation of the plasmonic networks. Furthermore, the DNMs extend efficient plasmon-assisted catalysis to an intermediate operation temperature window (225-300°C).
\par 
In conclusion, we demonstrate conversion of CO$_2$ into CO, CH$_4$ and C$_2$H$_6$ with Cu-Pd disordered network metamaterials at elevated temperatures. The reaction selectivity and catalytic yield can be tailored through the chemical engineering of the bimetallic Cu-Pd system. 
Probing the plasmonic environment on the nanoscale, we show that regions with a high local density of states coincide with high Pd content. In these regions light is localized in ultra small volumes. This combination of high local density of states and small mode volumes make our systems ideal for the plasmon assisted catalysis. Evidence of this is given by the changes in yield and selectivity of reaction products under light illumination. It is anticipated that the selectivity and therefore also the total production rate can further be enhanced by fine tuning of the chemistry of the plasmonic networks or substituting Pd with other catalytic metals such at Ru. Furthermore, measuring the quantum efficiency - how effectively a photocatalyst converts an absorbed photon into a product - could provide further insights into the underlying mechanisms.

\par 
\textbf{Materials and Methods} 
\par 
\textbf{Catalyst fabrication}
\par 
The thin film catalysts were fabricated in two steps. First, Al-Cu-Pd thin films (nominal 300~nm) were co-sputtered onto SiO$_2$/Si wafers ($381\pm25~\mu m$). Second, the films were chemically dealloyed in a 1 M NaOH aqueous solution, to form a disordered open-porous nano-network~\cite{wohlwend_chemicalengineering,wohlwend2023, wohlwend2024hybrid}. 
\par 
\textbf{Material characterisation}
\par 
Focused ion beam crosssections of the thin film catalyst networks were acquired using a NVision40 focused ion beam scanning electron microscope (FIB-SEM), operated at 3 kV and with a focused Ga$^+$ liquid metal ion source at 30 kV acceleration voltage. From the crosssection, the surface area of the nanonetwork as well as the weight were calculated by importing the crosssection into the open source software Blender and creating a 3D model (See Supporting Information). The reflectivity of the samples was measured using a deuterium / halogen lamp (Ocean Optics DH-2000-BAL) and a UV-Vis-NIR (188 - 1033nm wavelength) spectrometer (Ocean Optics USB2000+XR). As a calibration standard, an aluminium mirror was used.
\par 
EEL spectra of the thin film catalyst were obtained by a monochromated FEI Titan 60-300 with an imaging filter (Gatan GIF Quantum) operated in scanning mode at 300 kV. The spectra were acquired with a dispersion of 25 meV, respectively, 10 meV per pixel. All spectra were treated with the HQ Dark Correction plugin. Additionally, postprocessing of the EELS data included the alignment of the spectra with respect to the position of the zero-loss peak (ZLP), normalization of the maximum intensity of the ZLPs in each pixel and removal of the ZLP by fitting a premeasured ZLP using the Matlab Spectrum image analysis tool and DigitalMicrograph (Gatan)~\cite{wohlwend2023}.

\textbf{Photocatalytic activity test}\\ 
The photocatalytic conversion of CO$_2$ was carried out in the gas phase in a gas flow control system (Alicat Scientific), a custom-built heatable (max. 300 °C) photoreactor (Figure~\ref{fig:Catalysis_two}~\textbf{A}) equipped with different light sources (Thorlabs), and a gas chromatograph (GC) (Micro GC Fusion, Inficon) for gas analysis. Thereby, the GC is equipped with three modules (Rt-Molsieve 5A with Rt-Q-Bond Backflush, Rt-QBond, and Rt-U-Bond) and thermal conductivity detectors.
\par 
For catalytic testing, the thin film samples were loaded into the reactor chamber. The chamber was then purged with $N_2$. After the purging, a gas flux ( 9:1 $H_2$:CO$_2$ ) was introduced. The (photo-) thermal catalytic data was then recorded at different temperatures and for different network compositions. 

\newpage 
\textbf{Acknowledgements and Author contributions} \\
J.W. and H.G. conceived the research plan. G. H. and J. W. performed the EELS measurements. J.W.,  O.W., B.M. and S.K. fabricated the samples and performed catalytic measurements and their analysis. D.K. significantly contributed to the catalytic measurements and their analysis. J.W. and  O.W. performed sample characterizations, data analysis and visualization. J.W. wrote the manuscript. All authors reviewed and commented the manuscript. 
\par 
The catalytic measurements were performed in the laboratory for multifunctional materials of Prof. Markus Niederberger. Sputter deposition of thin films was performed at FIRST, the clean room facilities of ETH Zürich. The authors acknowledge the technical support of FIRST. EELS measurements were carried out at the Graz Centre for Electron Microscopy. The authors are very grateful to Till Kyburz for fabricating a custom reactor. The authors are also grateful to Sven Roediger.
\bibliography{Paper}

\end{document}